\documentclass[sigconf]{acmart}

\AtBeginDocument{%
  \providecommand\BibTeX{{%
    \normalfont B\kern-0.5em{\scshape i\kern-0.25em b}\kern-0.8em\TeX}}}

\copyrightyear{2022}
\acmYear{2022}
\setcopyright{rightsretained}
\acmConference[CHI '22]{CHI Conference on Human Factors in Computing Systems}{April 29-May 5, 2022}{New Orleans, LA, USA}
\acmBooktitle{CHI Conference on Human Factors in Computing Systems (CHI '22), April 29-May 5, 2022, New Orleans, LA, USA}
\acmDOI{10.1145/3491102.3501823}
\acmISBN{978-1-4503-9157-3/22/04}

\usepackage{xspace}
\usepackage{amsmath}

\usepackage{wrapfig}
\usepackage{enumitem}
\usepackage[noabbrev,capitalise,nameinlink]{cleveref}
\usepackage[utf8]{inputenc}

\newcommand{\system}{\textsc{Neo}\xspace}
\newcommand{\shelf}{interactive shelf\xspace}

\usepackage{xspace,xpunctuate}
\newcommand{\ie}{{i.e.,}\xspace}
\newcommand{\eg}{{e.g.,}\xspace}

\definecolor{JSONRed}{HTML}{EF5B54}
\definecolor{JSONBlue}{HTML}{328BEF}
\newcommand{\codeRed}[1]{{\texttt{\textcolor{JSONRed}{#1}}}\xspace}

\definecolor{urlBlue}{HTML}{007aff}

\newcommand{\actual}{X}
\newcommand{\predicted}{Y}

\begin{document}

\title{\system: Generalizing Confusion Matrix Visualization to Hierarchical and Multi-Output Labels}

\author{Jochen Görtler}
\authornote{Work done at Apple.}
\affiliation{%
  \institution{University of Konstanz}
  \city{Konstanz}
  \country{Germany}
}
\email{jochen.goertler@uni-konstanz.de}

\author{Fred Hohman}
\affiliation{%
  \institution{Apple}
  \city{Seattle}
  \state{WA}
  \country{USA}
}
\email{fredhohman@apple.com}

\author{Dominik Moritz}
\affiliation{%
  \institution{Apple}
  \city{Pittsburgh}
  \state{PA}
  \country{USA}
}
\email{domoritz@apple.com}

\author{Kanit Wongsuphasawat}
\affiliation{%
  \institution{Apple}
  \city{Seattle}
  \state{WA}
  \country{USA}
}
\email{kanitw@apple.com}

\author{Donghao Ren}
\affiliation{%
  \institution{Apple}
  \city{Seattle}
  \state{WA}
  \country{USA}
}
\email{donghao@apple.com}

\author{Rahul Nair}
\affiliation{%
  \institution{Apple}
  \city{Heidelberg}
  \country{Germany}
}
\email{rahul\_nair@apple.com}

\author{Marc Kirchner}
\affiliation{%
  \institution{Apple}
  \city{Heidelberg}
  \country{Germany}
}
\email{marc_kirchner@apple.com}

\author{Kayur Patel}
\affiliation{%
  \institution{Apple}
  \city{Seattle}
  \state{WA}
  \country{USA}
}
\email{kayur@apple.com}

\renewcommand{\shortauthors}{Görtler, et al.}

\begin{abstract}

%%%% 00-abstract.tex starts here %%%%

The confusion matrix, a ubiquitous visualization for helping people evaluate machine learning models, is a tabular layout that compares predicted class labels against actual class labels over all data instances.
We conduct formative research with machine learning practitioners at Apple and find that conventional confusion matrices do not support more complex data-structures found in modern-day applications, such as hierarchical and multi-output labels.
To express such variations of confusion matrices, we design an algebra that models confusion matrices as probability distributions.
Based on this algebra, we develop \system, a visual analytics system that enables practitioners to flexibly author and interact with hierarchical and multi-output confusion matrices, visualize derived metrics, renormalize confusions, and share matrix specifications.
Finally, we demonstrate \system's utility with three model evaluation scenarios that help people better understand model performance and reveal hidden confusions.

%%%% 00-abstract.tex ends here %%%%

\end{abstract}




%% 
%% Teaser figure 
\begin{teaserfigure}
  \includegraphics[width=\textwidth]{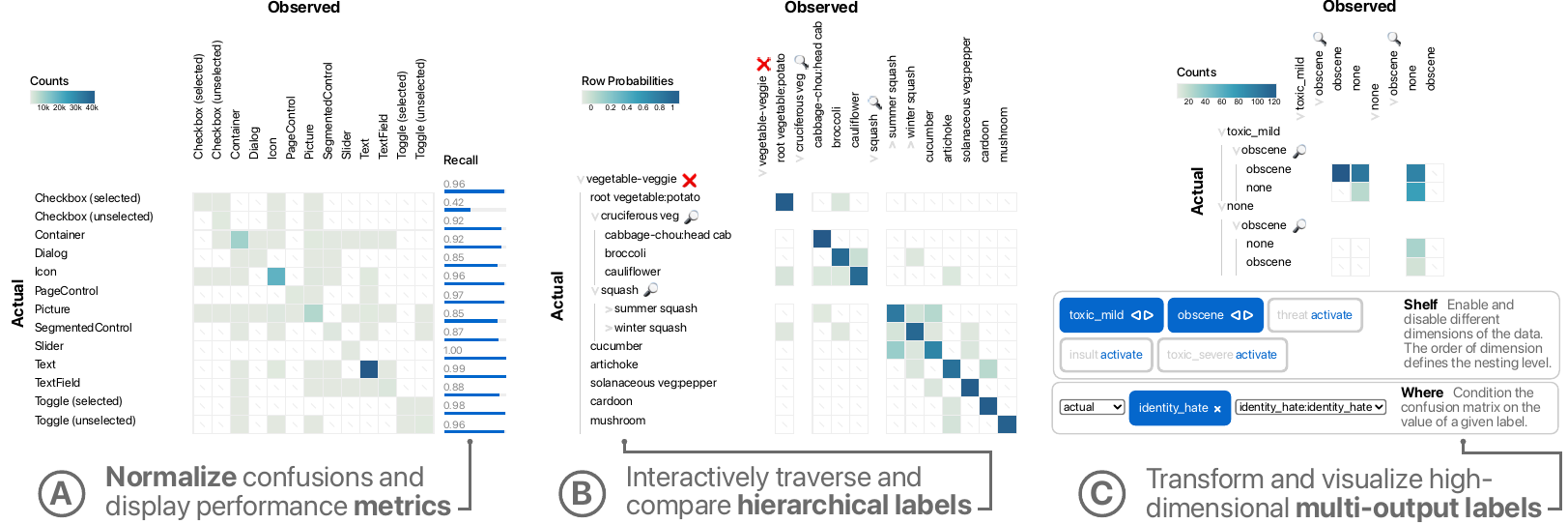}
  \caption{
    \system generalizes conventional confusion matrices and enables machine learning practitioners to find hidden confusions, visualize per class metrics, traverse hierarchical labels on tiered axes, and transform high-dimensional, multi-output labels for model evaluation.
    (\textbf{A}) This confusion matrix for an object detection model computes and shows user-specified performance metrics, such as recall, per class alongside the matrix visualization.
    (\textbf{B}) This sub-hierarchy of a confusion matrix for a 1,000-class image classifier compares the confusions of different vegetables, such as \textit{cucumber} and \textit{mushroom}, against the sub-hierarchies of \textit{cruciferous vegetables} and \textit{squash}.
    (\textbf{C}) This confusion matrix for a naive multi-output online toxicity detector conditions and filters by \textit{identity hate} confusions, nests \textit{obscene} confusions under \textit{mild toxic} confusions, and finds that the model misses to identify many \textit{obscene} comments.
  }
  \Description{
    Three different configurations of our new confusion matrix visualization system.
    (A) The first shows a bar chart of the recall metric next to the matrix visualization.
    (B) The second shows a matrix visualization with tiered class axes to traverse and compare hierarchical labels.
    (C) The third shows a toolbar where different classes are nested and filtered to transform and visualize high-dimensional multi-output labels.
  }
  \label{fig:teaser}
\end{teaserfigure}

\maketitle

%%%% 01-introduction.tex starts here %%%%

\section{Introduction}
\label{sec:introduction}

Machine learning is a complex, iterative design and development practice~\cite{patel2010gestalt,amershi2019software}, where the goal is to generate a learned model that generalizes to unseen data inputs.
One critical step is \textit{model evaluation}, testing and inspecting a model's performance on held-out test sets of data with known labels.
Due to the size of modern-day machine learning applications,
interactive data visualization has been shown to be an invaluable tool to help people understand model performance~\cite{amershi2015modeltracker,ren2016squares,wexler2017facets,hinterreiter2020confusionflow}.

A ubiquitous visualization used for model evaluation, particularly for classification models, is the \textit{confusion matrix}: a tabular layout that compares a predicted class label against the actual class label for each class over all data instances.
In a typical configuration, rows of the confusion matrix represent actual class labels and the columns represent predicted class labels (synonymously, these can be flipped via a matrix transpose).
These visualizations are introduced in many machine learning courses and are simultaneously used in practice to show what pairs of classes a model confuses.
Succinctly, confusion matrices are the ``go-to'' visualization for classification model evaluation.

Despite their ubiquity, conventional confusion matrices suffer from multiple usability concerns.
Confusion matrices show a visual proxy for accuracy (\eg entries on the diagonal of the matrix), which alone has been shown to be insufficient for many evaluations~\cite{yang2018grounding}.
Furthermore, the diagonal of a confusion matrix often contains many more instances than off-diagonal entries (can be orders of magnitude), which hides important confusions (\ie off-diagonal entries).
As practitioners improve their model, the net effect moves off-diagonal instances to the diagonal, further exacerbating this problem of hiding confusions.
Ironically, the better the model optimization, the harder it is to find confusions.
Confusion matrices also suffer from scalability concerns, \eg when a dataset has many classes, has strong class imbalance, has hierarchical structure, or has multiple outputs.

We believe confusion matrices can be significantly improved to help practitioners better evaluate their models.
To understand specific challenges around using confusion matrices, we conducted a formative research study through a survey with machine learning practitioners at Apple.
We found that in many machine learning applications confusion matrices become cumbersome to use at scale, do not show other metrics model practitioners need to know (\eg precision, recall), and can be hard to share.
Moreover, confusion matrices only support flat, single-label data structures; more complex yet common hierarchical labels and multi-output labels are not supported.

Informed by findings from the formative research and a literature review, we create a \textit{confusion matrix algebra}, which models confusion matrices as probability distributions and produces a unified solution for pitfalls of conventional confusion matrices.
Based on this algebra, we design and develop \system, a visual analytics system that enables practitioners to flexibly author and interact with confusion matrices in diverse configurations with more complex label structures.
The design of \system extends the confusion matrix, allowing users to visualize additional metrics for analysis context, inspect model confusions interactively through multiple normalization schemes, visualize hierarchical and multi-output labels, and easily share confusion matrix configurations with others.
\system maintains the familiar format of confusion matrices, using a conventional confusion matrix as the basis of the visualization.

In this work, our contributions include:

\begin{itemize}
    \item \textbf{Formative research}, including common challenges and analysis tasks, from surveying machine learning practitioners at Apple about how confusion matrices and model evaluation visualizations are used in practice.
    
    \item \textbf{A confusion matrix algebra} that generalizes and models confusion matrices as probability distributions.
    
    \item \textbf{\system},\footnote{\url{https://github.com/apple/ml-hierarchical-confusion-matrix}} \textbf{a visual analytics system} for authoring and interacting with confusion matrices that supports hierarchical and multi-output labels.
    \system also introduces a specification (or colloquially, a ``spec'') that enables sharing specific visualizations with others.
    \system is reactive in that authoring a spec updates the visualization, and interacting with the visualization updates the spec.
    
    \item \textbf{Three model evaluation scenarios} demonstrating how \system helps practitioners evaluate machine learning models across domains and modeling tasks, including object detection, large-scale image classification, and multi-output online toxicity detection.
\end{itemize}

We believe machine learning should benefit everyone.
Understanding where models fail helps us correct them to enable better experiences.
We hope the lessons learned from this work inform the future of model evaluation while inspiring deeper engagement from the burgeoning intersection of human-computer interaction and artificial intelligence.

%%%% 01-introduction.tex ends here %%%%

%%%% 02-related-work.tex starts here %%%%

\section{Related Work}
\label{sec:related-work}

\label{subsec:related-work-performance}
Model evaluation is a key step to successfully applying machine learning.
However, what it means for a model to perform well greatly depends on the task. 
A variety of metrics have been developed to evaluate classifiers~\cite{hossin2015review}; common example metrics include \textit{accuracy}, \textit{precision}, and \textit{recall}.
However, there is no one-size-fits-all metric,\footnote{There is no better illustration of this than viewing the overwhelming number of different metrics that can computed from a confusion matrix: \url{https://en.wikipedia.org/wiki/Confusion_matrix\#Table_of_confusion}.} and the utility of metrics depend on the modeling task.

\paragraph{Model Performance Visualizations}\label{subsec:related-work-vis}
The visualization community has developed novel visual encodings to help practitioners better understand their model's performance.
These techniques can be categorized as either \textit{class-based} or \textit{instance-based}.
For class-based techniques, 
\citet{alsallakh2014visual} use a radial graph layout where the links represent confusion between classes.
\citet{seifert2009novel} embed all test and training samples into a radial coordinate systems where units are classes.
Regarding instance-based visualization,
\citet{amershi2015modeltracker} propose a unit visualization that shows how each instance is classified and shows the closeness of instances in the feature space.
\emph{Squares}~\cite{ren2016squares} extends this visualization and shows, per class, how instances are classified within a multi-classifier.
Similarly, \emph{ActiVis}~\cite{kahng2017activis} uses instance-based prediction results to explore neuron activations in deep neural networks.
While we focus on practitioners, previous work has also studied confusion matrix literacy and designed alternative representations for non-expert and public algorithmic performance understanding~\cite{shen2020designing}.
Common to these works is that they introduce new visualization concepts that may not be familiar to machine learning practitioners and therefore require training and adaption time.

\paragraph{Confusion Matrix Visualizations}
Instead of introducing alternative visual encodings, our approach aims to enhance confusion matrices directly, a ubiquitous visualization that already has familiarity within the machine learning community~\cite{wilkinson2009history}, and adapt them to types of data that are encountered in practice today.
There exists some work that enhances conventional confusion matrices.
For example, individual instances have been shown directly in the cells of a confusion matrix~\cite{wexler2017facets, bruckner2014mloscope}.
\citet{bilal2017convolutional} investigate hierarchical structure in neural networks using confusion matrices.
In their work, hierarchies can be constructed interactively based on blocks in the confusion matrix, which are then shown using icicle plots.
They also provide group level statistics for the elements of the hierarchy.
However, in contrast to our work, their system does not consider multi-output labels.

Confusion matrices are also used in iterative model improvement.
\citet{hinterreiter2020confusionflow} propose a system to track confusions and model performance over time by juxtaposing confusion matrices of different modeling runs.
Their system also provides an interactive shelf to specify the individual runs.
Our work also features an \shelf; however, its purpose in our work is to drill down into sub-hierarchies of a larger confusion matrix.
Furthermore, confusion matrices have been used to directly interact with machine learning models.
As such, they can be used to interactively adapt decision tree classifiers~\cite{elzen2011baobab}, by augmenting them with information about the splits that are performed by each node.
For models that are based on boosting, \citet{talbot2009ensemble} propose a system to adjust the weights of weak classifiers in an ensemble to achieve better performance.
Furthermore, \citet{kapoor2010interactive} propose a technique to interactively steer the optimization of a machine learning classifier, based on directly interacting with confusion matrices.
None of these systems consider hierarchical or multi-output labels.

There are also different approaches that generalize beyond conventional confusion matrices.
Class-based similarities from prediction scores instead of regular confusions have been proposed to generalize better to hierarchical and multi-output labels~\cite{alsallakh2020visualizing}.
\citet{zhang2019manifold} embed pairwise class prediction scores into a Cartesian coordinate system to compare the performance of different models.
Furthermore, multi-dimensional scaling has been used to embed confusion matrices into 2D~\cite{susmaga2004confusion}.
In contrast to our work, these adaptions stray further from familiar confusion matrix representations.

\paragraph{Model Confusions as Probability Distributions}
Framing the confusion matrix as a probability distribution has been used by the machine learning community to investigate classifier variability~\cite{calaen2017bayesian}.
In addition, other work shows how a probabilistic view of the confusion matrix can be used to quantify classifier uncertainty~\cite{toetsch2020classifier}.
However, both these works only consider binary classification.
Preliminary work on generalizing to multi-label problems~\cite{krstinic2020multi} computes the contribution of an instance to a cell, but only if the prediction was only partially correct.
Our work builds upon these views and produces a unified language for generalizing confusion matrices to hierarchical and multi-output labels.

\paragraph{Table Algebra}\label{subesc:related-work-algebra}
Our work is inspired by \textit{relational algebra theory}~\cite{codd1970relational} and the \textit{table algebra} in \textit{Polaris}~\cite{stolte2002polaris}, now the popular software \textit{Tableau}, and its work on visualizing hierarchically structured data~\cite{stolte2002query}.
In Polaris, a user can visually explore the contents of a database by dragging variables of interest onto ``shelves''.
The contents of a shelf are then transformed into queries to a relational database or OLAP cube, which retrieves the data for visualization.
Our approach is different in that we support operations on matrices and that it is based on probability distributions rather than a relational database model.

%%%% 02-related-work.tex ends here %%%%

%%%% 03-survey.tex starts here %%%%

\section{Formative Research: Survey, Challenges, \& Tasks}
\label{sec:survey}

To understand how practitioners use confusion matrices in their own work, we conducted a survey that resulted in 20 responses from machine learning researchers, engineers, and software developers focusing on classification tasks at Apple.
Respondents were recruited using an internal mailing list about machine learning tooling and targeted practitioners who regularly use confusion matrices.
We take inspiration from the methods used in previous visualization literature on multi-class model visualization~\cite{ren2016squares}, bootstrapping our survey questions from their work.
The survey consists of eight questions centered around machine learning model evaluation and confusion matrix utility.
The first two questions (\textbf{Q1} and \textbf{Q2}) are multiple choice, while the remaining (\textbf{Q3–Q6}) are open responses.

\begin{itemize}

    \item[\textbf{Q1.}] Which stages of machine learning do you typically work on?
    \item[\textbf{Q2.}] How many classes does your data typically have?
    \item[\textbf{Q3.}] When do you use confusion matrices in your ML workflow?
    \item[\textbf{Q4.}] Which insights do you gain from using confusion matrices?
    \item[\textbf{Q5.}] Which insights are missing, or you wish you would also gain from using a confusion matrix?
    \item[\textbf{Q6.}] How often are your labels structured hierarchically (for example \textit{apple} could be in the category \textit{fruit}, which then is part of the category \textit{food})? How do you work with hierarchical confusions? How deep are the hierarchies?
    \item[\textbf{Q7.}] When do you encounter data where one instance has multiple labels (for example an instance that is \textit{apple} and \textit{ripe})? How many labels are typically associated with an instance?
    \item[\textbf{Q8.}] How else do you visualize your data and errors besides confusion matrices? What are their advantages?
\end{itemize}

From the survey data, we used thematic analysis to group common purposes, practices, and challenges of model evaluation into categories~\cite{gibbs2007thematic}.
Throughout the discussion, we use representative quotes from the respondents to illustrate the main findings.

\begin{figure}
    \centering
    \includegraphics[width=1.0\columnwidth]{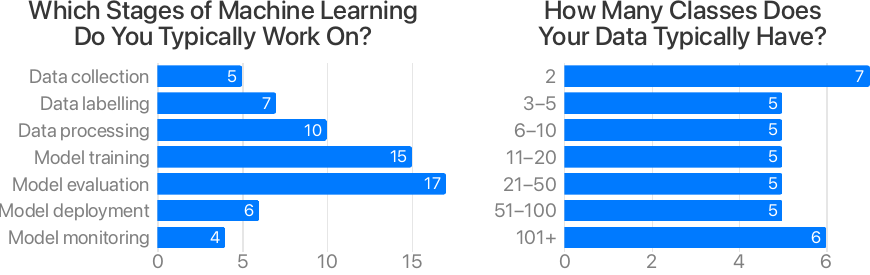}
    
    \caption{
        Survey responses from machine learning practitioners (multiple choice questions).
        Left: respondents cover every stage of machine learning process; many of them work on ``data processing'' and ``model training,'' with the majority of respondents indicating ``model evaluation,'' the specific machine learning stage we focus on in this work.
        Right: respondents work on classification models of a variety of sizes, ranging from binary classifiers to models with over 1,000 classes.
    }
    \Description{
      Two charts visualizing survey responses from machine learning practitioners (multiple choice questions).
      The first question asks ``Which stages of machine learning do you typically work on?'' and highlights the most responses for ``model evaluation''.
      The second question asks ``How many classes does your data typically have?'' and shows that most responses are spread, but most models have under 50 class.
    }
    \label{fig:survey}
\end{figure}

\subsection{Respondents' Machine Learning Backgrounds}

% (Q1, Q2)

We asked practitioners what stages of the machine learning process they typically work on (\textbf{Q1}, multiple choice), to establish context about the respondents' backgrounds.
The left-hand side of \cref{fig:survey} shows a histogram of what stages our respondent's have experience in, sorted by the chronological order of stages in the machine learning development process~\cite{amershi2019software}.
With some expertise represented at every stage of machine learning development, most respondents indicate their work falls between ``data processing,'' ``model training,'' and ``model evaluation.''
This diverse experience and concentration on model evaluation give us, the researchers, confidence that the population of practitioners surveyed contains the relevant experience and knowledge to speak about the intricacies of model evaluation with confusion matrices.

To gain insight into the scale of the respondents' modeling work, we also asked about the typical number of classes modeled from their datasets (\textbf{Q2}, multiple choice).
The right-hand side of \cref{fig:survey} shows a histogram for these responses, sorted from the fewest number of classes (binary classification) to the largest (101+). 
Results show an emphasis on binary classification, a majority skew towards models with fewer than 50 classes, but also representation from larger-scale models with over 100 classes.
These results establish that our respondents have worked with small-scale datasets, large-scale datasets, and everything in-between, strengthening our confidence that many different machine learning applications are represented in our formative research.

\subsection{Why Use Confusion Matrices?}

% (Q3, Q4)

We first categorize and describe the reasons why respondents use confusion matrices (\textbf{Q3}).
While we expected certain use cases to be reported, we were surprised by the number of roles and responsibilities confusion matrices satisfy in practice, such as performance analysis, model and data debugging, reporting and sharing, and data annotation efforts (\textbf{Q4}).

\subsubsection{Model Evaluation and Performance Analysis}

Confusion matrices are constructed to evaluate, test, and inspect class performance in models; therefore, it is unsurprising that most of the responses, 14/20, indicate that model evaluation is the main motivation for using confusion matrices in their own work.
Respondents explain that detailed model evaluation is critical to ensure machine learning systems and products produce high-accuracy predictions or a good user experience.
According to one respondent, confusion matrices allow a practitioner to see \textit{``performance at a glance.''}
One frequent and primary example reported was checking the presence of a strong diagonal; diagonal cells indicate correctly predicted data instances (whereas cells outside the diagonal represent confusions), therefore a strong diagonal is found in well-performing models.

\subsubsection{Debugging Model Behavior by Finding Error Patterns}

Besides seeing performance at a glance, 7/20 respondents indicated that confusion matrices are also useful for identifying error patterns to help debug a model.
Regarding pattern identification, a respondent said confusion matrices \textit{``allow me to see how a certain class is being misclassified, or if there is a pattern in misclassifications that can reveal something about the behavior of my model.''}
Respondents described multiple common patterns practitioners look for, including checking the aforementioned strong diagonal of the matrix, finding classes with the most confusions, and finding classes that are over-predicted.
Another interesting pattern reported by a natural-language processing practitioner was determining the directionality of confusions for a pair of classes.
For example, in the case of a bidirectional language translation model, does a particular sentence correctly translate from the source to the target language, but not the reverse.
These patterns can be \textit{``...much more revealing than a simple number,''} and help practitioners find shared similarity between two confused classes.

\subsubsection{Communication, Reporting, and Sharing Performance}

While confusion matrices help an individual practitioner understand their own model's behavior, they are also used in larger machine learning projects with many invested stakeholders.
Here, it is critical that team members are aware of the latest performance of a model during development, or monitoring the status of a previously-deployed model that is evaluated on new data.
One respondent reported that \textit{``exporting the matrix is more useful,''} since confusion matrices are commonly shared in communication reports with other individuals.

\subsubsection{High-quality Data Annotation}

Beyond model evaluation, respondents said confusion matrices are also useful for data labelling/annotation work.
Machine learning models require large datasets to better generalize, which results in substantial efforts to obtain high-quality labels.
In this use case, a practitioner wants to understand annotation performance instead of model performance; in some scenarios the same practitioner fulfills both roles.

Some newly labelled datasets undergo quality assurance, where a subset of the newly labelled data is scrutinized, adjusted, and corrected if any labels were incorrectly applied.
These labels are then compared against the original labelled dataset using a confusion matrix.
These data label confusion matrices visualize the performance of a data annotator (could be human or computational) instead of a model's performance (which can also be thought of as an annotator).
This process allows practitioners to find data label discrepancies between different teams.
For example, one respondent reported they \textit{``get to understand if there are certain labels or prompts that are causing confusion between the production [label] team and the quality assurance [label] team.''}
The quality assurance team often shares these visualizations with the production labeling teams to \textit{``improve the next [labeling] iteration,''} guiding annotation 
efforts through rich and iterative feedback.

\subsection{Challenges with Confusion Matrices}

% (Q5)

When prompted about where confusion matrices may not be sufficient (\textbf{Q5}), respondents voiced that they have experienced challenges (\textbf{C1–C4}) due to limitations with its representation and lack of support for handling more complex dataset structure. 
Visualizations for these datasets either did not exist or were shoehorned into existing confusion matrices by neglecting or abusing label names and structure (\textbf{Q8}).

\subsubsection{Hidden Performance Metrics (\textbf{C1})}

% (Q8)

The most common limitation of conventional confusion matrices discovered from our survey is their inability to show performance metrics for analysis context.
Over half, 11/20, respondents said that it was important to see other metrics alongside confusions (\textbf{Q8}).
Even accuracy is not explicitly listed in a confusion matrix but must be computed from specific cells for each class, which can be taxing when performing the mental math over and over.
While respondents listed other important performance metrics such as precision, recall, and true/false positive/negative rates, deciding which metrics are important is specific to the modeling task and domain.
Lastly, when sharing confusion matrices with others, respondents said it is important to provide textual descriptions of performance to help focus attention on specific errors.

\subsubsection{Complex Dataset Structure: Hierarchical Labels (\textbf{C2})}

% (Q6)

Another big challenge for confusion matrices is capturing and visualizing complex data structures that are now common in machine learning applications.
Conventional confusion matrices assume a flat, one-dimensional structure, but many datasets today across data types have hierarchical structure.
When asked specifically about dataset structure, 9/20 respondents said they work with hierarchical data and that typical model evaluation tools, like confusion matrices, do not suffice (\textbf{Q6}).
For example, an \textit{apple} class could be considered a subset of \textit{fruit} which is a subset of \textit{food}.
One respondent indicated that their team works almost exclusively with hierarchical data.
In the applications with hierarchical data, respondents indicated that the hierarchies were on average 2–4 levels deep (\ie from the root node to the leaf nodes).
Handling hierarchical classification data and the subgroups inherent to its structure is currently not supported in confusion matrix representations

\subsubsection{Complex Dataset Structure: Multi-output Labels (\textbf{C3})}

% (Q7)

Another type of dataset structure complexity is also well-represented in our survey, namely datasets that have instances with multi-output labels (\textbf{Q7}).
For example, an \textit{apple} could be \textit{red} and \textit{ripe}.
Over half, 11/20, of the respondents indicated that they work with datasets with multi-output labels that conventional confusion matrices do not support.
In such datasets, respondents said that, on average, data instances have 1–3 labels each, but one respondent described an application where instances had 20+ labels.
It is important to also note the distinction between labels and metadata: metadata is any auxiliary data describing an instance, whereas a label denotes a specific model output for prediction.
In short, all labels are metadata, but not all metadata are labels.

\subsubsection{Communicating Confusions while Collaborating (\textbf{C4})}

We have already identified and discussed the need for communicating model performance and common confusions in collaborative machine learning projects.
However, there remains friction when sharing new model results with confusion matrices, for example, a loss of quality and project context (\eg copying and pasting charts as images into a report).
It can be time consuming for a practitioner to prepare and polish a visualization to include in a report, yet it is important to ensure model evaluation is accessible to others.
Some respondents said it would be convenient if their confusion matrices could be easily exported.
This challenge is twofold: what are better and sensible defaults for confusion matrix visualization, and how can systems reduce the friction for practitioners sharing their latest model evaluations?

\subsection{Motivation and Task Analysis}
\label{subsec:tasks}

From our formative research, there is clear opportunity to improve confusion matrix visualization.
Practitioners reported that conventional confusion matrices, while useful, are insufficient for many of the recent advancements and applications of machine learning, and expressed enthusiasm for visualization to better help understand model confusions.
This research also yielded several key ideas that inspired us to rethink authoring and interacting with confusion matrices.
To inform our design, we distill tasks that practitioners perform to understand model confusions.
Tasks (\textbf{T1–T4}) map one-to-one to challenges (\textbf{C1–C4}):

\begin{itemize}

    \item[\textbf{T1}.] Visualize derived performance metrics while enabling flexible data analysis, such as scaling and normalization (\textbf{C1}).
    
    \item[\textbf{T2}.] Traverse and visualize hierarchical labels (\textbf{C2}).
    
    \item[\textbf{T3}.] Transform and visualize multi-output labels (\textbf{C3}).
    
    \item[\textbf{T4}.] Share confusion matrix analysis and configurations (\textbf{C4}).
\end{itemize}

%%%% 03-survey.tex ends here %%%%

%%%% 04-grammar.tex starts here %%%%

\begin{figure*}[t!]
    \centering
    \includegraphics[width=\textwidth]{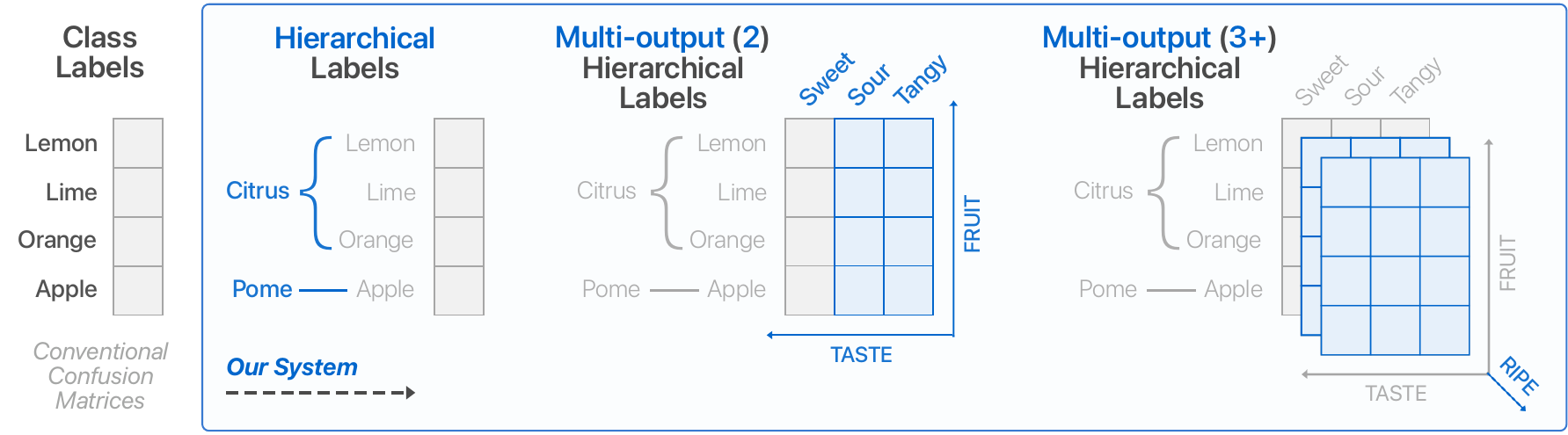}
    \caption{
        A visual representation of class labels for conventional confusion matrices (left) compared to our work (in blue) that supports \textbf{hierarchical labels} and \textbf{multi-output labels}.
        To build a confusion matrix from any of these label structures, compute every combination of the actual label against the predicted label for all classes.
    }
    \Description{
      A diagram showing how conventional confusion matrices only support flat, one-dimensional label structures, whereas ours can support hierarchical and multi-output labels.
      The diagram builds on the conventional label structure (a single array) by adding new dimensions represented as matrices.
    }
    \label{fig:algebra}
\end{figure*}

\section{Confusion Matrix Algebra}
\label{sec:algebra}

From our formative research, we aim to generalize confusion matrices to include hierarchical and multi-output labels.
For these types of data, analysis usually requires data wrangling as a preprocessing step, where practitioners develop one-off scripts.
Our work takes a different approach:
We provide a unified view of the different operations and analysis tasks for confusion matrices in the form of a specification language (\textbf{T4}) that is based on a key insight: \textbf{\textit{Confusion matrices can be understood as probability distributions}}.
While this way of viewing confusion matrices may seem unwieldy at first, its expressiveness becomes clear when we think about how practitioners interact with hierarchical and multi-output labels (\cref{fig:algebra}).

Confusion matrices show the number of occurrences for each combination of an actual class versus a predicted class.
Rows in a confusion matrix represent actual classes, columns represent predicted classes, and the cells represent the frequencies of all combinations of actual and predicted classes.
Our algebra leverages that the actual class $\actual$ and the predicted class $\predicted$ can be viewed as variables in a multivariate probability distribution $P(\actual,\predicted)$.
The probability mass function of this distribution is given by the relative frequencies of occurrences, which we obtain by dividing the absolute frequencies by the number of instances in the dataset.
For an introduction to multivariate probabilities, we recommend the book by \citet{hogg2020probability}.
Here we will explain the concepts of our algebra using the labels $\textit{Fruit} = \{ \text{apple}, \text{orange}, \text{lemon} \}$  as an example.
In this setting, the following describes a cell in the confusion matrix, specifically \textit{apples} that are mistaken for \textit{oranges}:
$$
P(\text{Fruit}_{\actual}=\text{apple} \,,\, \text{Fruit}_{\predicted}=\text{orange}).
$$

This probabilistic framing allows us to use the standard operations of multivariate probability distributions to transform our data.
In particular, we use the following operations, which we also illustrate in \cref{fig:multiclass}:
\textbf{Conditioning} primes a probability distribution on given values.
We can use this operation to extract sub-views of a larger confusion matrix.
\textbf{Marginalization} allows us to discard variables of multivariate distributions that we are currently not interested in by summing over all such variables.
These operations have the algebraic property that their results are again probability distributions---mathematically this is defined as \textit{closedness}.
This property is not purely theoretical, but rather it also has practical implications:
It allows us to chain multiple operations together to form complex queries.
Moreover, the algebra automatically ensures correct normalization after every step.
In addition to the two operations above, we also propose a \textbf{nesting} operation, which is useful to investigate multiple labels simultaneously.

\subsection{Normalization}
\label{sec:algebra-normalization}

Normalization is essential for confusion matrices as it determines how the data is visualized (\textbf{T1}).
Our probabilistic framework guarantees normalization implicitly, as all objects are probability distributions.
Depending on the task, it might make sense to normalize a confusion matrix by rows or columns.
Choosing a normalization scheme can emphasize patterns that large matrix entries might otherwise hide (example shown in \cref{subsec:case-study-1}).
Normalizing by rows or by columns also produces \textit{recall} and \textit{precision}, two widely used performance metrics echoed from our formative research.
The recall for a label is the value on the diagonal of a matrix normalized by rows: $P(\text{Fruit}_{\predicted} = \text{orange} \,|\, \text{Fruit}_{\actual} = \text{orange})$.
Similarly, the precision for a label is the value on the diagonal of a matrix normalized by columns:
$P(\text{Fruit}_{\actual} = \text{orange} \,|\, \text{Fruit}_{\predicted} = \text{orange})$.
Both cases can be computed using Bayes' rule.

\subsection{Hierarchical Labels}
\label{sec:algebra-hierarchical}

With our algebra, practitioners can understand how confusions relate to hierarchical labels by drilling down into specific sub-hierarchies (\textbf{T2}).
In addition, we can use the hierarchical structure to improve the visual representation of large confusion matrices by collapsing sub-hierarchies.
Collapsing sub-hierarchies is equivalent to summarizing multiple entries.
First, we collect all the rows/columns that belong to the category to be collapsed.
In terms of probability distributions, we create a compound probability (here for \textit{Citrus}) for these items:
\begin{gather*}
P(\text{Fruit}_{\actual} = \text{Citrus} \,,\, \text{Fruit}_{\predicted} = \text{Citrus}) = \nonumber \\
P(\text{Fruit}_{\actual} \in \{ \text{lemon}, \text{orange} \} \,,\, \text{Fruit}_{\predicted} \in \{ \text{lemon}, \text{orange} \} ) \nonumber
\end{gather*}
This rewrite is possible because, for visualization, the individual rows/columns of a confusion matrix are not affected by another---they are \textit{mutually independent}.
Therefore, we can conclude $P(\text{Citrus}) = P(\text{lemon}) + P(\text{orange})$.

The other type of analysis that our algebra supports is drilling down into a sub-hierarchy.
For this, we will condition the multivariate distribution on the rows/columns that we want to consider:
\begin{align}
P(\text{Fruit}_{\actual} \,,\, \text{Fruit}_{\predicted} \,|\, \text{Fruit}_{\actual} = \text{Citrus} \,,\, \text{Fruit}_{\predicted} = \text{Citrus}) \nonumber
\end{align}
This operation results in a new confusion matrix that only contains the specified rows and columns as shown in \cref{fig:multiclass}.

\subsection{Multi-output Labels}
\label{sec:algebra-multiclass}

Multi-output labels make it significantly harder to evaluate a model's performance.
The number of cells in a confusion matrix grows exponentially for datasets with multi-output labels.
Adding an additional label $\textit{Taste} = \{ \text{sweet}, \text{sour}, \text{tangy} \}$ to the fruit dataset results in $81$ possible combinations of actual and predicted states: $$
|\text{Fruit}_{\actual}| \times |\text{Fruit}_{\predicted}| \times |\text{Taste}_{\actual}| \times |\text{Taste}_{\predicted}| = 3 \times 3 \times 3 \times 3 = 81$$

\begin{figure}
  \centering
  \includegraphics[width=1.0\columnwidth]{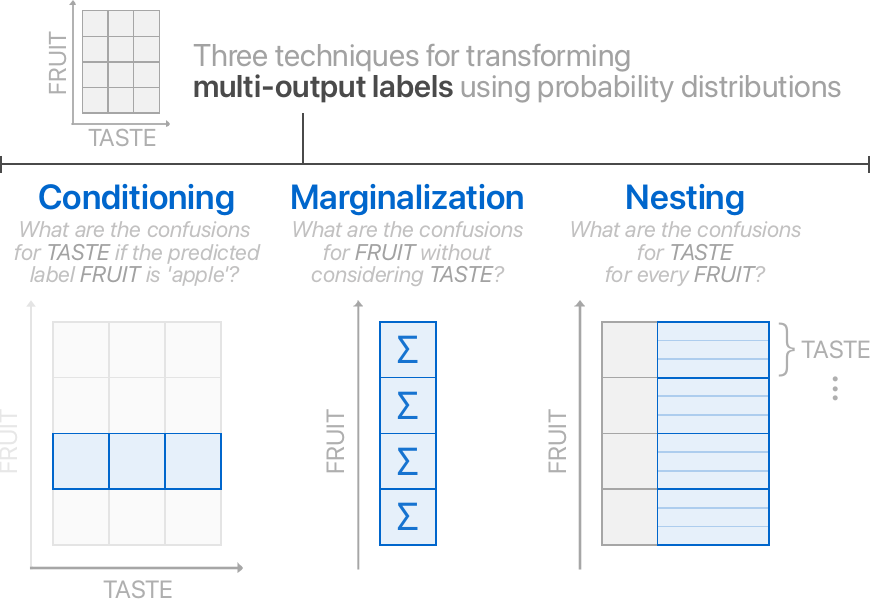}
  \caption{
      A visual representation of the three techniques for transforming high-dimensional multi-output labels.
      First, we can \textbf{condition} the confusion matrix based on the value of another label.
      To focus on a single label, we can use \textbf{marginalization} to sum across ignored labels.
      We can also \textbf{nest} multiple labels to form hierarchical labels.
  }
  \Description{
    A diagram showing the three techniques to transform high-dimensional labels. 
    From a two-dimensional matrix from ``fruit'' and ``taste'' dimensions, (1) conditioning pulls out a row (all ``tastes'' for one fruit) of the matrix, (2) marginalization sums along the rows (``fruits'') of a matrix, and (3) nesting builds a table of ``tastes'' per ``fruit''.
  }
  \label{fig:multiclass}
\end{figure}

Our algebra provides multiple techniques to transform high-dimensional confusions into 2D for different analyses (\textbf{T3}), illustrated in \cref{fig:multiclass}.
In the following discussion, we use example analysis questions to ground the explanation of each technique.

Initially, an analyst might ask \textit{What are the confusions for ``Taste'', if the predicted label was ``apple''?}, \ie we consider confusions for one label given a class of a different label.
We achieve this in our algebra by conditioning the multivariate distribution on this class.
The following example shows the confusion matrix only for \textit{apples}:
\begin{align}
P(\text{Taste}_{\actual}\,,\, \text{Taste}_{\predicted}\,|\, \text{Fruit}_{\actual} = \text{apple}\,,\, \text{Fruit}_{\predicted} = \text{apple}) \nonumber
\end{align}
This operation usually changes the number of columns and rows of the resulting confusion matrix because not all labels necessarily occur together with the fixed label (\cref{fig:multiclass}, left).

Furthermore, an analyst may currently not be interested in one of the variables and ask: \textit{What are the confusions for ``Fruits'' without considering their ``Taste''?}
In this case, we can discard the needless variable in our probabilistic framework using marginalization.
Here, we discard \textit{Taste}:
\begin{gather*}
P_{\text{Fruit}}(\text{Fruit}_{\actual}\,,\, \text{Fruit}_{\predicted}) =\\
\sum_i \sum_j P(\text{Fruit}_{\actual}\,,\, \text{Fruit}_{\predicted}\,,\, \text{Taste}_{\actual}(i) \,,\, \text{Taste}_{\predicted}(j)) \nonumber
\end{gather*}
Note that this operation does not change the dimensionality of the variables that we are interested in but instead sums over the frequencies of the discarded entries accordingly (\cref{fig:multiclass}, middle).

Finally, analysts that need to understand the relationship between two different variables may ask: \textit{What are the confusions for the ``Taste'' for every ``Fruit''?}
To inspect multiple dimensions simultaneously, our algebra can nest one label below another.
Multiple labels in a dataset form a high-dimensional confusion matrix, which cannot be readily visualized using a 2D matrix representation.
The nesting operation solves this problem by realizing all possible combinations of labels in a structured manner (the \textit{power set} of the variables) and induces a hierarchical structure---the relationship between parent and child is given by the ordering of the nesting (\cref{fig:multiclass}, right).
This is a useful technique for visualizing joint distributions.

%%%% 04-grammar.tex ends here %%%%

%%%% 05-system.tex starts here %%%%

\section{\system: Interactive Confusion Matrix Visualization}
\label{sec:system}

To put our confusion matrix algebra into practice, we design and develop \system, a visual analytics system that enables practitioners to flexibly author and interact with confusion matrices for model evaluation.
Our visualization system is agnostic to the model architecture and data. 
If the classification problem (or data annotation task) can record instance labels and predictions, \system can ingest the results.
Throughout the following section, we link relevant views and features to the tasks (\textbf{T1–T4}) identified from our formative research (\cref{subsec:tasks}).

\subsection{Design Goal: Preserve Familiar Confusion Matrix Representation}

Whereas many machine learning visualizations do not have an established form, confusion matrices have an expected and borderline ``standardized'' representation.
Instead of reinventing the confusion matrix visualization, our primary design goal for \system was to leverage the familiarity of confusion matrices and improve upon their functionality with complementary views and interaction.
For example, in the simplest case where a practitioner has a classification model with a dataset whose instances have no hierarchy and only one class label, \system shows a conventional confusion matrix.
However, even in these cases there is still opportunity for improving model evaluation through interaction.

\begin{figure}[t]
    \centering
    \includegraphics[width=1.0\columnwidth]{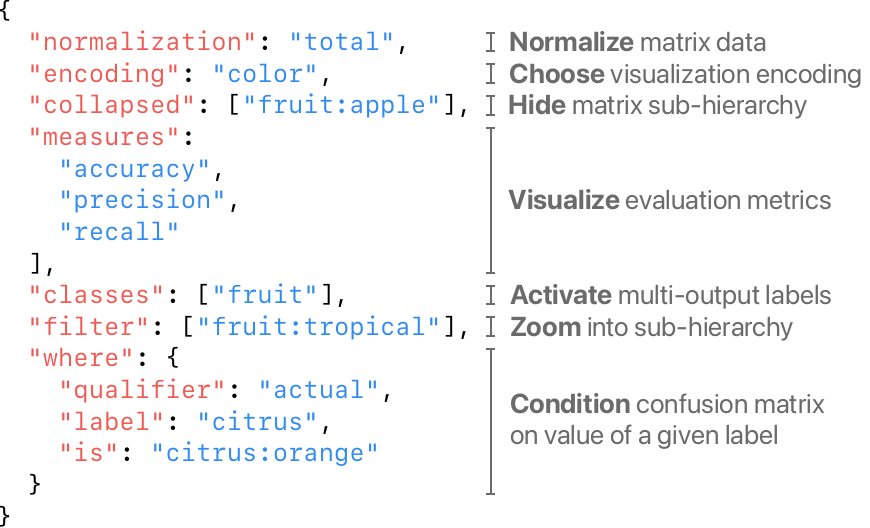}
    \caption{
        \system's JSON specification based on our confusion matrix algebra.
        The specification configures the confusion matrix based on the selected \codeRed{normalization} scheme, visualization \codeRed{encoding}, and desired \codeRed{measures}, but also saves the shown state of the hierarchy (\codeRed{collapsed}, \codeRed{filter}) and multi-output labels using either marginalization (\codeRed{classes}), nesting (order of \codeRed{classes} array), or conditioning (\codeRed{where}).
    }
    \Description{
      A JSON code snippet of the matrix specification colored by different fields.
    }
    \label{fig:spec}
\end{figure}

\subsection{Specification for Matrix Configuration}

\system is built upon a powerful domain-specific language (DSL) for specifying a confusion matrix configuration.
Implemented in \system using JSON, this paradigm provides similar benefits to other declarative specification visualizations~\cite{satyanarayan2016vega}: automated analysis, reproducibility, and portability.
\cref{fig:spec} shows an example ``spec'' and its different fields.
In this section we describe every field of the spec.

\system is a \textit{reactive system}: configuring the spec updates the visualization, and interacting with the visualization updates the spec.
This is a powerful interaction paradigm where a practitioner can tailor their desired view using either code or the interface while remaining in sync~\cite{kery2020mage}.
Once a practitioner is satisfied with their visualization, they can easily share their spec with others since their view is represented as a JSON string (\textbf{T4}).
In \system, the spec is hidden by default, but is exposed through a single button click.

\subsection{Interacting with Confusion Matrices}

The primary view of \system is the confusion matrix itself (multiple examples seen in \cref{fig:teaser}).
Rows represent actual classes and columns represent predicted classes.
A cell contains the number of data instances incorrectly predicted from the row class as the column class; the exception are cells along the diagonal that indicate the number of correctly predicted instances for a particular class.
To see how many instances are in each cell, one can hover over any cell to display a textual description of the confusion count.
This feature was requested in our formative research.
Moreover, hovering over a cell highlights its row and column in the matrix, using a light amber background color (\cref{fig:encoding}A), to ease a user's eye-tracking when reading the axis labels.
We chose to use a visual encoding for confusion counts instead of numeric labels within each cell, since with bigger data (e.g., cells with confusions larger than 3 digits) the labels become long, grow the size of each cell, and ultimately inhibit the number of classes can fit in one display.
Furthermore, for models with many classes, from our formative research practitioners wanted to a high-level overview of the performance of a model first with the option to inspect specific cells, hence the design of the details-on-demand textual descriptions.

\subsubsection{Visualization Encodings and Confusion Normalization}

\begin{figure}[t!]
    \centering
    \includegraphics[width=1.0\columnwidth]{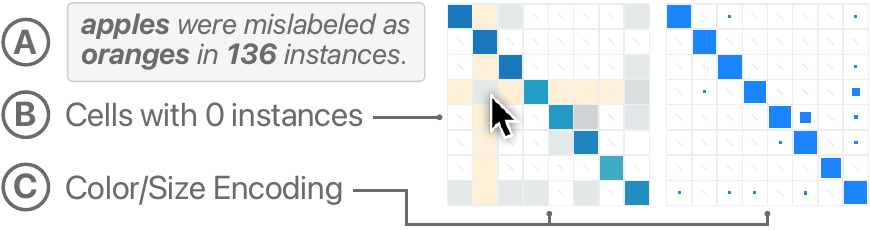}
    \caption{
        (\textbf{A}) Brushing a cell in \system displays the confusion information in a natural language caption.
        (\textbf{B}) Confusion matrix cells with a 0 value are excluded from the encoding scale.
        (\textbf{C}) Users can choose between a color and size encoding for the confusion matrix cells.
    }
    \Description{
      Two small confusion matrices visualized using color and size to show the differences between the encodings.
      A mouse pointer hovers over a cell to display a tooltip of how many instances belong to that cell.
    }
    \label{fig:encoding}
\end{figure}

The default encoding is \texttt{color} (arguably the default in practice).
Users can toggle between a \texttt{color} encoding and a \texttt{size} encoding where inner squares are scaled to support comparison of absolute values (\cref{fig:encoding}C); this is set in the specification in the \codeRed{encoding} field (\cref{fig:spec}).
Regardless of encoding, this common representation already presents a problem with confusion matrices: the diagonal contains many more instances than off-diagonal entries (\eg orders of magnitude), which hide important confusions in the matrix.
As practitioners improve a model over time, the net outcome moves instances from off-diagonal entries to the diagonal, further exacerbating this problem.
Ironically, the better the model optimization, the harder it is to see confusions.

\system addresses this issue in multiple ways.
First, \system leverages a color discontinuity for the value 0~\cite{kandel2012profiler}.
Cells with 0 instances are not colored and instead contain a small light-gray dash, which makes it immediately clear which cells have confusions and which do not (\cref{fig:encoding}B).
Second, \system can scale the color of the matrix by everything except the diagonal, giving the full color range exclusively to the confusions (the diagonal is removed from the visualization in this case).
Third, practitioners can choose from different normalization schemes, presented in detail in \cref{sec:algebra-normalization}, to see different views of the confusions (\textbf{T1}).
The default normalization scales cells by the instance count, but \system supports normalizing by the \texttt{rows} or \texttt{columns}.
We can read recall and precision respectively from the diagonal of the normalized matrix.
Normalization is set in the spec in the \codeRed{normalization} field (\cref{fig:spec}).

\subsubsection{Performance Metrics Per Class}

Related to choosing different normalization schemes, respondents from our formative research indicated that confusion matrices lack analysis context for looking at other metrics alongside the visualization.
Performance metrics, such as accuracy, precision, recall, and others are not readily accessible from a confusion matrix.
Aggregate metrics such as these can also be broken down from the model-level to the class-level to support better class-by-class analysis.
\system solves this problem by visualizing both aggregate and per-class metrics on the right-hand side of the confusion matrix as an additional column per metric (\cref{fig:teaser}A and \cref{fig:case-study-1}), where the top number corresponds to the aggregate metric, and numbers aligned with each row correspond to each class.
Besides the metrics listed above, \system also includes metrics such as the count of actual and predicted instances, true/false positives, and true/false negatives.
These are all set in the spec in the \codeRed{measures} field (\cref{fig:spec}).
While this addition may seem small, it is one of the most common limitations of conventional confusion matrices and was 
continuously requested by respondents from our formative research.

\subsection{Visualizing Hierarchical Labels}

Hierarchical datasets are one of the more complex structures discovered from our formative research that conventional confusion matrices do not support (see \cref{sec:algebra-hierarchical}).
Following our design goal to preserve the confusion matrix representation, \system supports hierarchical labels (see \cref{fig:algebra}) through multiple design improvements (\textbf{T2}).
First, the class labels on the axes are nested according to the hierarchy, where classes further in the hierarchy are indented (see \cref{fig:teaser}B and \cref{fig:case-study-2}).
Second, the matrix is partitioned into blocks based on the lowest hierarchy level.
Hovering over any cell in the matrix highlights its parent hierarchy indicator black (vertical gray bars) for easier tracking (\cref{fig:case-study-2}A).
Together, these two improvements help users understand model performance with the hierarchy directly represented in the visualization.

\system interactively collapses sub-hierarchies in two ways (see \cref{fig:case-study-2}).
First, selecting a parent class on either axis toggles between showing or hiding the children classes.
The interaction collapses (or expands) the parent class and recomputes the confusion data to accurately represent the new aggregate class category in the matrix (\textbf{T2}).
This implements a Focus+Context paradigm, by expanding class categories of interest while keeping surrounding categories available nearby~\cite{card1999readings}.
\system models hierarchies as virtual category trees~\cite{sun2001hierarchical}, and expands and collapses sub-matrices symmetrically, since the asymmetric case makes confusion much harder to reason about.
Alternatively, selecting the magnifying glass icon triggers a drill-down, replacing the entire visualization with only the selected sub-hierarchy and remaps the color (or size) encoding.
These techniques allow practitioners to explore larger confusion matrices by reducing the number of visible classes shown and comparing class categories against one another.
Regardless of technique, the spec is also updated to record which sub-hierarchies are collapsed or zoomed, set in the \codeRed{collapsed} and \codeRed{filter} fields respectively (\cref{fig:spec}), ensuring that when returning to \system in the future, or sharing the current view, a user picks up where they left off (\textbf{T4}).

\subsection{Visualizing Multi-output Labels}

\begin{figure}
  \centering
  \includegraphics[width=1.0\columnwidth]{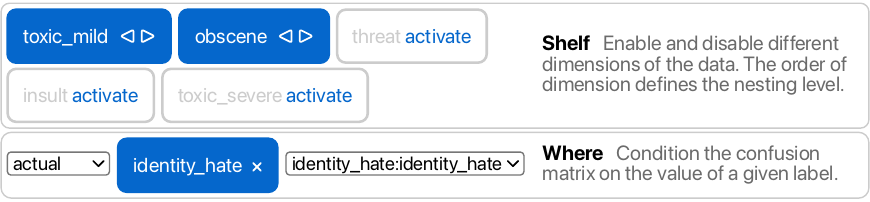}
  \caption{
      \system's \shelf let's practitioners specify how to transform multi-output labels for visualization.
      Non-activated (gray) multi-output labels are \textit{marginalized}.
      Activated (blue) multi-output labels define a \textit{nesting} order.
      The confusion matrix can also be \textit{conditioned} on the value of a particular label.
  }
  \Description{
    A screenshot from the system's interactive shelf.
    It shows an example combination of selecting a label to remove it from marginalization, nesting labels by arranging their selected order, and conditioning labels using a drop down menu.
  }
  \label{fig:shelf}
\end{figure}

Multi-output labels are another more complex structure discovered from our formative research unsupported by conventional confusion matrices (see \cref{sec:algebra-multiclass}).
Analyzing multi-output models is difficult since confusions are represented in an unbounded high-dimensional space (see \cref{fig:algebra}), which inhibits directly applying conventional matrix visualization.
To preserve the confusion matrix representation familiarity, \system supports three mechanisms to transform high-dimensional confusions into 2D (\textbf{T3}): conditioning, marginalization, and nesting (for details of each, see \cref{sec:algebra-multiclass}).
Inspired by previous work in exploratory visualization~\cite{stolte2002polaris, wongsuphasawat2017voyager}, in \system, visualizing multi-output labels leverages an \textit{\shelf} to specify label transformations (\cref{fig:shelf}).
The \shelf contains all multi-output labels for a given dataset.
Multi-output labels are either activated or not; activating a multi-output label toggles its color from blue to gray.
Activating a multi-output label displays the label in the confusion matrix for analysis. 

\paragraph{Conditioning}

The first technique to transform multi-output confusions is conditioning, \ie analyzing confusions for one label given a class of a different label.
In these scenarios, \system conditions the confusion matrix based on the value of a specified label.
A practitioner can select to condition the matrix on an actual or predicted class from the conditioning label in the \shelf.
Note that when a multi-output label is used for conditioning, it can no longer be used for nesting.
Similar to the other techniques, these options are reflected in the spec in the \codeRed{where} field (\cref{fig:spec}).

\paragraph{Marginalization}

To visualize high-dimensional confusions, another technique uses marginalization to sum over all other multi-output labels that a practitioner is not interested in.
Therefore, in the \shelf in \system, multi-output labels that are not activated, \ie grayed out instead of blue, are marginalized automatically.
In the spec, activated classes are kept in-sync and saved in the \codeRed{classes} field (\cref{fig:spec}).

\paragraph{Nesting}

Oftentimes a practitioner wants to inspect several multi-output labels at once.
To address this issue, \system nests multi-output labels under one another. 
Nesting multi-output labels creates a hierarchical label structure, which \system already supports, where each class of the child label is replicated across all classes of the parent label.
\system automatically nests multi-output labels when more than one label is activated in the \shelf.
Reordering the labels in the shelf changes the nesting order.
This order is also reflected in the spec as the order of the activated classes in \codeRed{classes} field (\cref{fig:spec}).

\subsection{System Design and Implementation}

\system is a modern web-based system built with Svelte\footnote{Svelte: \url{https://svelte.dev}}, TypeScript\footnote{TypeScript: \url{https://www.typescriptlang.org}},
and D3\footnote{D3: \url{https://d3js.org}}.
The spec is implemented as a portable JSON format to easily share confusion matrix configurations with other stakeholders (\textbf{T4}).
Regarding system scalability, \system is bounded by conventional SVG constraints in the browser (e.g., displaying tens of thousands of SVG elements).
Engineering effort such as leveraging Canvas or WebGL would remove this constraint, however we believe better interactions for configuring confusion matrices to compare relevant classes and submatrices is more helpful to practitioners than rendering the biggest matrix possible.

%%%% 05-system.tex ends here %%%%

%%%% 06-case-studies.tex starts here %%%%

\begin{figure*}
  \centering
  \includegraphics[width=\textwidth]{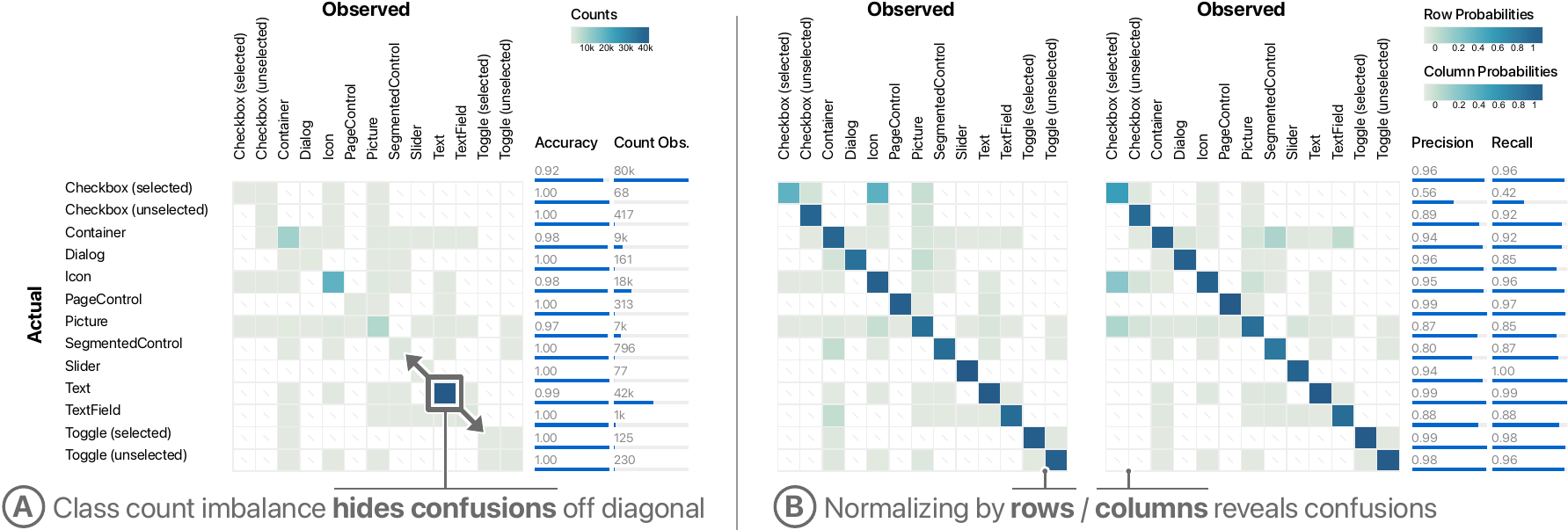}
  \caption{
      (\textbf{A}) When a dataset has class count imbalance (\eg some class have many more data instances than others, as seen by the ``Count Obs.'' metric), confusions off the diagonal are hidden and the ``Accuracy'' metric is misleading.
      (\textbf{B}) Normalizing by row and/or column probabilities reveals hidden confusions, and has direct connections to other more appropriate model evaluation metrics such as precision and recall.
  }
  \Description{
    A confusion matrix whose dataset has class imbalance, which is reflected in the metrics bar chart in the system.
    A user then normalizes the matrix to reveal hidden confusions (e.g., the matrix cells change in color).
  }
  \label{fig:case-study-1}
\end{figure*}

\section{Model Evaluation Scenarios}
\label{sec:case-studies}

The following three model evaluation scenarios showcase how \system helps practitioners evaluate models across different domains, including object detection (\cref{subsec:case-study-1}), large-scale image classification (\cref{subsec:case-study-2}), and online toxicity detection (\cref{subsec:case-study-3}).

\subsection{Finding Hidden Confusions}
\label{subsec:case-study-1}

Recent work on screen recognition showed how machine learning can create accessibility metadata for mobile applications directly from pixels~\cite{zhang2021screen}.
An object detection model trained on 77,637 screens extracts user-interface elements from screenshots on-device.
The publication includes a confusion matrix for a 13-class classifier that reports and summarizes model performance (test set contains 5,002 instances).
With \system, we can further analyze this confusion matrix and find hidden confusions to help improve the end-user experience of the model.

First, \system loads the confusion matrix with the default ``Accuracy'' metric appended on the right-hand side, as seen in \cref{fig:case-study-1}A.
By excluding cells with 0 confusions from the visualization, we can quickly see which class pairs have confusions and which do not.
Looking at the accuracies, we see good performance across classes, but in the visualization notice that a few cells dominate the color encoding.
When a dataset has strong class count imbalance, \eg class distribution is not equal, ``Accuracy'' is a misleading metric to use for evaluating a multi-class model.
We confirm this by adding the ``Count Observed'' metric in the specification to see that the \textit{Text} and \textit{Icon} classes contain many more instances, 42k and 18k respectively (\cref{fig:case-study-1}A).

With \system, we normalize the confusion matrix by the row or column probabilities, seen in \cref{fig:case-study-1}B, that automatically remap the color encoding to reveal hidden confusions.
These normalizations are closely related to two other metrics, precision and recall, practitioners use to better inspect performance per class.
After adding these metrics to the spec, we see low recall (with row normalization) and low precision (with column normalization) for the \textit{Checkbox (selected)} class; digging into the confusion matrix shows errors with the \textit{Icon} class (\cref{fig:case-study-1}B, right).
We also see confusions between the \textit{Container} class and \textit{SegmentedControl} and \textit{TextField} that were previously hidden in \cref{fig:case-study-1}A.
\system's design, metrics, and normalization features make error analysis actionable by surfacing hidden error patterns to model builders.

\begin{figure*}[t!]
  \centering
  \includegraphics[width=\textwidth]{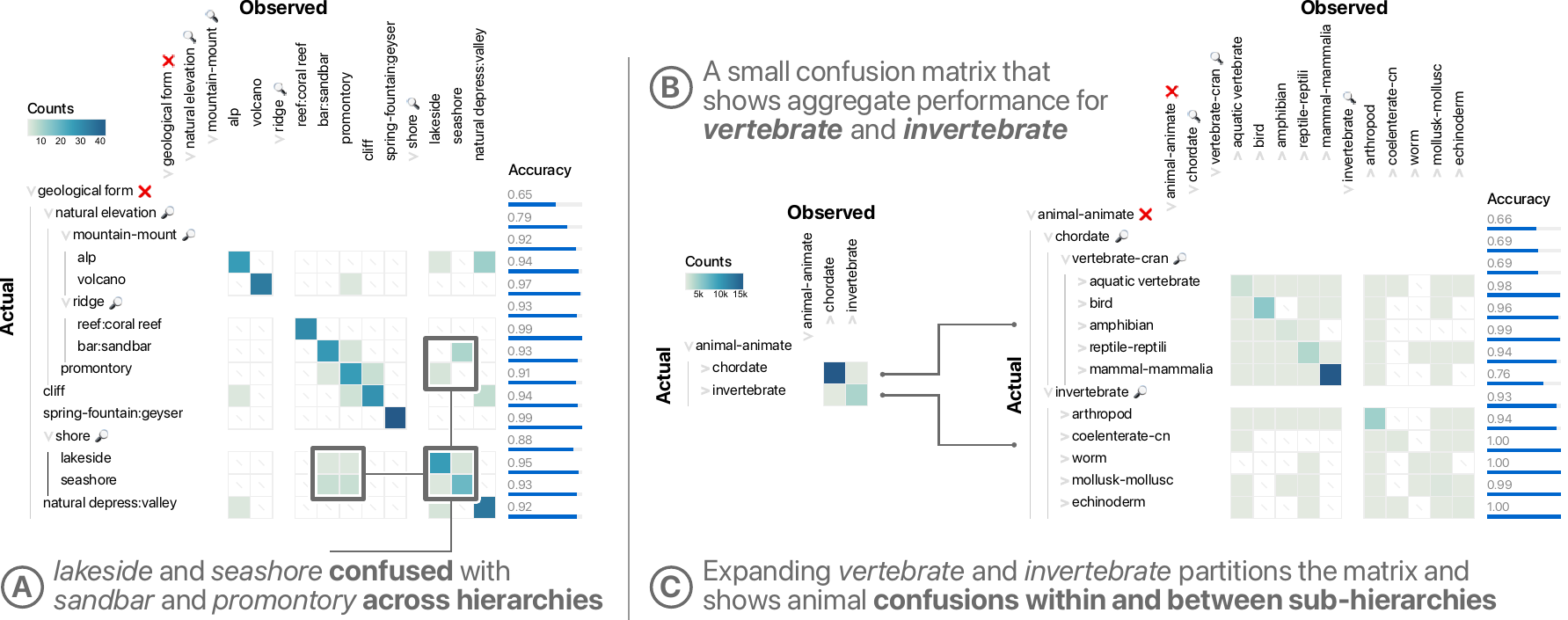}
  \caption{
      (\textbf{A}) In a deep learning model trained on ImageNet, \system reveals the \textit{geological form} sub-hierarchy contains confusions between semantically related classes across sub-hierarchies.
      (\textbf{B}) Another high-level sub-hierarchy for \textit{animal-animate} expands (\textbf{C}) to show detailed confusion comparisons within and between sub-hierarchies of animal classes.
  }
  \Description{
    A confusion matrix that highlights blocked confusions between different groups of labels from natural landmarks.
    A user then expands and collapses the hierarchical labels on the matrix axis to inspect the grouped confusions.
  }
  \label{fig:case-study-2}
\end{figure*}

\subsection{Traversing Large, Hierarchical Image Classifications}
\label{subsec:case-study-2}

Achieving high accuracy on ImageNet, with its 1.2M+ data instances spread across 1,000 classes, is a standard large-scale benchmark for image classification.
Most work considers ImageNet classes as a flat array, but the classes originate from the WordNet~\cite{miller1995wordnet} hierarchy.
To test \system's scalability, we analyze the results of a ResNet152-V2~\cite{he2016identity} deep learning model trained on ImageNet, including its hierarchical structure.
The validation set contains 50,000 images.

When loading a large hierarchical confusion matrix, \system defaults to collapsing all sub-hierarchies and starting at the root.
In this configuration, the metrics show the aggregate performance of the entire model, but as we expand into sub-hierarchies these metrics are recomputed per sub-hierarchy and class.
Beginning at the root node of the hierarchy, we expand to an early sub-hierarchy titled \textit{object-physical} that contains three sub-categories, each of which we filter for analysis.
The first category, \textit{part-portion}, expands fully to contain classes of cloth and towels.
The performance on this sub-hierarchy is rather good (strong diagonal), so we continue.
Second, the \textit{geological-form} category expands fully to contain classes of natural landscapes (\cref{fig:case-study-2}A).
While the accuracy is high on most classes (above 91-99\%), one sub-hierarchy, \textit{shore}, is lower than the others (88\%).
\textit{Shore} contains two classes, \textit{lakeside} and \textit{seaside}.
There are a few confusions between the two, which is expected given their semantic similarity, but there exists another set of confusions between these classes and \textit{sandbar} and \textit{promontory} (point of high land that juts out into a large body of water), which both belong to a different sub-hierarchy (\cref{fig:case-study-2}A).
\system enables practitioners to discover these confusions across different sub-hierarchies.

The third and final category, \textit{whole-unit}, in our original sub-hierarchy contains hundreds of classes.
We are now interested in inspecting the performance of living things in our model, \ie the \textit{organism-being} category, which contains four sub-hierarchies: \textit{person-individual}, \textit{plant-flora}, and \textit{fungus} all perform well, but \textit{animal-animate} contains many classes with confusions.
Expanding \textit{animal-animate} shows two classes of interest: \textit{chordate-vertebrate} and \textit{invertebrate}, a biological distinction between groups of animals that do or do not have a backbone.
This 2x2 confusion matrix is useful for comparing this meaningful, high-level sub-hierarchy (\cref{fig:case-study-2}B), but expanding both categories one level deeper presents multiple directions for deeper analysis by comparing confusions from animal classes within and between one another (\cref{fig:case-study-2}C).
Throughout this analysis, \system's specification automatically updates the configuration, so the exact view can be saved and shared with any other project stakeholder.

\begin{figure*}[t!]
  \centering
  \includegraphics[width=\textwidth]{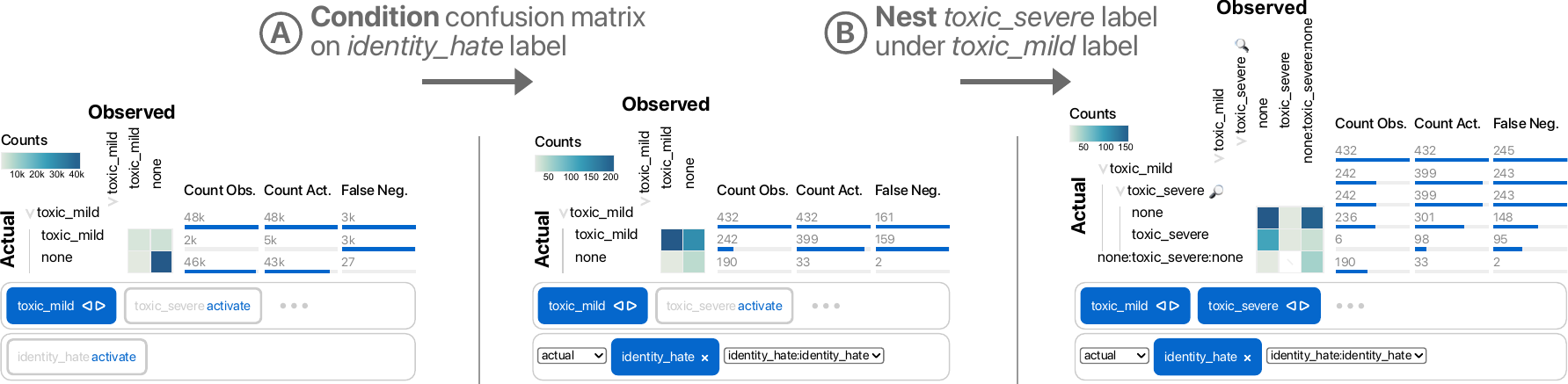}
  \caption{
      (\textbf{A}) Conditioning a confusion matrix for a toxicity classification model on \textit{identity hate} comments filters all confusions according to the value of the label.
      (\textbf{B}) Nesting the confusion matrix by \textit{toxic mild} and \textit{toxic severe} allows us to visualize both labels simultaneously. 
  }
  \Description{
    A small confusion matrix that changes its color when a user conditions the data, then changes in size as a user nests one label under another. 
  }
  \label{fig:case-study-3}
\end{figure*}

\enlargethispage{12pt}

\subsection{Detecting Multi-class and Multi-label Online Toxicity}
\label{subsec:case-study-3}

To make online discussion more productive and respectful, an open challenge tasked developers to build toxicity classifiers~\cite{kaggle2017toxic}.
From 159,571 Wikipedia comments labeled by people for toxic behavior, this is a multi-class, multi-label classification problem containing 6 types of non-mutually exclusive behavior: \textit{mild toxicity}, \textit{severe toxicity}, \textit{obscene}, \textit{threat}, \textit{insult}, and \textit{identity hate}; \eg a comment can be both \textit{mildly toxic} and a \textit{threat}.
We analyze the results from a naive one-vs-rest logistic regression classifier from a test set of 47,872 comments.

\system defaults to loading \textit{mild toxic} comments as the first label to consider.
Because this is a multi-output label confusion matrix, \system visualizes the 2x2 matrix of \textit{mild toxic} comments against \textit{none}, \ie everything else.
The \shelf tells us that the other 5 classes are currently marginalized.
In \cref{fig:case-study-3}, left, the ``Count Obs.'' metric tells us this dataset has a large class imbalance, \ie there are many more non-toxic comments than there are toxic comments. 
This means our model could struggle with false negatives.
Checking this metric indicates that indeed, of the approximately 5k \textit{mild toxic} comments, our naive model only correctly predicts around 2k of them, leaving nearly 3k false negatives.
This is an early indication that our model architecture may not suffice for this dataset.

We are interested in other hurtful discussion that could cause emotional harm, therefore we only want to consider \textit{identity hate} comments.
To do this, we condition the confusion matrix on \textit{identity hate} (\cref{fig:case-study-3}A).
\cref{fig:case-study-3}, middle, shows that the model is better at identifying \textit{mild toxic} comments given the instances are also \textit{identity hate}, but there are still \textit{mild toxic} false negatives present.

\enlargethispage{12pt}

Beyond \textit{mild toxic}, we now want to inspect more serious comments within \textit{severe toxic}.
In \system, we activate and nest \textit{severe toxic} comments under \textit{mild toxic} comments to consider the occurrence of both multi-output labels simultaneously (\cref{fig:case-study-3}B).
From \cref{fig:case-study-3}, right, we see the model correctly identifies some of these comments, but suffers a similar problem as \textit{mild toxic} comments in that it has many false negatives.
We could consider other confusion matrix configurations, such as nesting \textit{obscene} under \textit{mild toxic} comments to form a larger hierarchy as shown in \cref{fig:teaser}C, but already, we can confidently conclude that our first model cannot distinguish between \textit{mild toxic} comments and benign comments.
To improve this model, the next step is likely choosing a different architecture, such as a long short-term memory model, that can learn richer features from the raw text data.

%%%% 06-case-studies.tex ends here %%%%

%%%% 07-discussion.tex starts here %%%%

\section{Discussion and Future Work}
\label{sec:discussion}

Our work opens up many research directions that envision a future where confusion matrices can be further transformed into a powerful and flexible performance analysis tool.

\paragraph{Confusion Matrix Visualization Scalability}

Scaling confusion matrix visualization remains an important challenge.
In \system, hierarchical data can be collapsed to focus on smaller submatrices, since most of the time comparing classes within nearby categories is more meaningful, \eg comparing ``apple'' to ``orange'' instead of ``apple'' to ``airplane.''
However, in the case of a one-dimensional, large confusion matrix (\eg more than 100+ classes), the conventional representation suffers from scalability problems.
Beyond scrolling and zooming, we envision richer interactions and extensions to our algebra to handle larger confusion matrices.
For example, investigating a ``NOT'' operation that ignores columns to produce a better color mapping to find smaller matrix cells, or leveraging classic table seriation techniques~\cite{bertin1983semiology}.

\paragraph{Automatic Submatrix Discovery}

Related to scalability, further algorithmic advancements could help automatically find interesting submatrices of a confusion matrix.
From our formative research, this was briefly discussed in the context of a large matrix.
Automatically finding groups of cells based on some metric, \eg low-performing classes, could help guide a practitioner towards important confusions to fix and reduce the number of cells on screen.

\paragraph{Interactive Analysis with Metadata}

In data annotation efforts, other metadata is collected besides the raw data features and label(s).
How can we use this other metadata to explore model confusions?
For example, in an image labelling task, annotators may be asked to draw a bounding box around an object.
We could then ask questions about patterns of confusions when metadata such as ``bounding box'' was small, represented in our confusion matrix algebra as:
$P(X, Y \,|\,\text{bounding box area} < A)$.
We could also compute new metrics, such as the percentage of small bounding boxes when the ``apple'' class was confused with ``orange,'' represented in our confusion matrix algebra as: 
$P(\text{bounding box area} < A \,|\, X = \text{apple} \,,\, Y = \text{orange})$.
Interactive query interfaces that support these types of questions could help practitioners attribute confusions to specific features or metadata, saving time when searching for error patterns.

\paragraph{Comparing Model Confusions Over Time}

Machine learning development is an inherently iterative process~\cite{hohman2020understanding,amershi2019software,patel2010gestalt,hinterreiter2020confusionflow}, where multiple models are often compared against each other.
Two common comparison scenarios include (1) training multiple models at once, and (2) retraining a model after debugging.
In the first scenario, it would be useful to interactively compare confusion matrices against one another to select the best performing model.
In the second scenario, using a confusion matrix to compare an improved model against its original version could help practitioners test whether their improvements were successful.
One potential comparison technique could be to take the difference between confusion matrices to find the biggest changes.

\paragraph{Creating Datasets from Confusions}

While visualizing confusion matrices and aggregate errors helps practitioners debug their models, it can be useful to inspect individual data instances.
From our formative research, practitioners expressed interest in extracting instances from confusion matrix cells.
Future interactions such as filtering by confusion type and previewing instances within each cell could support extracting and creating new subsets of data for future error analysis.

%%%% 07-discussion.tex ends here %%%%

%%%% 08-conclusion.tex starts here %%%%

\section{Conclusion}
\label{sec:conclusion}

From our formative research, one respondent reported that \textit{``confusion matrices are one type of analysis when analyzing performance... doing thorough analysis requires looking at lots of different distributions of the data.''}
This quote raises a keen point that while confusion matrices remain a ubiquitous visualization for model evaluation, they are only one view into model behavior.
There is no one-size-fits-all model evaluation visualization, nor one magic model metric.
Regardless, confusion matrices continue to be an excellent tool to teach modeling fundamentals to novices and an invaluable tool for practitioners building industry-scale systems.

In this work, we generalize the capabilities of confusion matrices while maintaining their familiar representation.
Through formative research, we design an algebra that models confusion matrices as probability distributions and expresses more variations of confusion matrices, e.g., datasets with hierarchical and multi-labels.
Based on this algebra, we develop \system, a visual analytics system that allows practitioners to flexibly author, interact with, and share confusion matrices.
Lastly, we demonstrate \system's utility with three model evaluation scenarios that help people better understand model performance and reveal hidden confusions.

%%%% 08-conclusion.tex ends here %%%%

\begin{acks}

%%%% 00-acknowledgments.tex starts here %%%%

We thank our colleagues at Apple for their time and effort integrating our research with their work.
We especially thank Lorenz Kern who helped with initial prototyping datasets.
Jochen Görtler is supported in part by the Deutsche Forschungsgemeinschaft (DFG, German Research Foundation) – Project-ID 251654672 – TRR 161.

%%%% 00-acknowledgments.tex ends here %%%%

\end{acks}

\balance
\bibliographystyle{ACM-Reference-Format}
%%%% bibliography starts here %%%%

\bibliography{22-neo-chi}%% Commented by merge tool

% \appendix

\end{document}